\documentclass[bibyear]{aa} 

\usepackage{natbib}
\usepackage{graphicx}
\usepackage{txfonts}
\usepackage{hyperref}
\usepackage{float}
\usepackage{ulem}
\usepackage{xcolor}
\usepackage{subcaption}
\usepackage[switch]{lineno}

\bibpunct{(}{)}{;}{a}{}{,} 

\usepackage[para]{threeparttable}
\usepackage{dcolumn}    
\newcolumntype{.}{D{.}{.}{-1}}
\newcolumntype{0}{D{.}{.}{0}}

\begin{document} 
   \title{ Tracing satellite planes in the Sculptor group:}
 \subtitle{II. Discovery of five faint dwarf galaxies in the DESI Legacy Survey}  
   
   \titlerunning{Discovery of six dwarf galaxies}
    \authorrunning{Mart{\'\i}nez-Delgado et al.}


   \author{David Mart{\'\i}nez-Delgado\inst{1,2,3}\thanks{ARAID Fellow}, Michael Stein$^{4}$, Marcel S. Pawlowski$^{5}$, Joanna D. Sakowska\inst{3,6}, Dmitry Makarov$^{7}$, Lidia Makarova$^{7}$, Giuseppe Donatiello$^{8}$, Dustin Lang$^{9}$}
   
\institute{
$^{1}$ Centro de Estudios de F\'isica del Cosmos de Arag\'on (CEFCA), Unidad Asociada al CSIC, Plaza San Juan 1, 44001 Teruel, Spain\\
$^{2}$ ARAID Foundation, Avda. de Ranillas, 1-D, E-50018 Zaragoza, Spain\\
$^{3}$ Instituto de Astrofísica de Andalucía, CSIC, Glorieta de la Astronom\'\i a,  E-18080 Granada, Spain\\
$^{4}$ Ruhr University Bochum, Faculty of Physics and Astronomy, Astronomical Institute (AIRUB), 44780 Bochum, Germany \\
$^{5}$ Leibniz-Institut f\"ur Astrophysik Potsdam (AIP), An der Sternwarte 16, D-14482 Potsdam, Germany\\
$^{6}$ Department of Physics, University of Surrey, Guildford GU2 7XH, UK \\
$^{7}$ Special Astrophysical Observatory of the Russian Academy of Sciences, Nizhnij Arkhyz, 369167, Russia \\
$^{8}$UAI -- Unione Astrofili Italiani /P.I. Sezione Nazionale di Ricerca Profondo Cielo, 72024 Oria, Italy \\
$^{9}$Perimeter Institute for Theoretical Physics, 31 Caroline St N, Waterloo, Canada\\
}
\date{Received XXXX; accepted XXXX}


  \abstract
  {Although substantial progress has been made in reconciling $\Lambda$CDM simulations with the observed abundance and distribution of satellite galaxies, important tensions persist. Studying satellite systems around spiral galaxies thus remains key in addressing these tensions.}
  {In this series of papers, we report the first results of an on-going systematic survey of faint dwarf spheroidal galaxies in the vicinity of the bright late-type spiral NGC\,253 galaxy,  the brightest member of the Sculptor filament located at a distance of 3.7~Mpc towards Anti-Virgo, using Dark Energy Camera images.}
   {We performed a new NGC\,253 satellite search by means of visual inspection using co-added image cutouts reprocessed in the DESI Legacy image surveys, reaching a very low surface brightness regime (28.0--29.0~mag\,arcsec$^{-2}$).
   We used \textsc{galfit} software for deriving the photometric and structural properties of the five dwarf galaxy candidates.  }
  {Five new dwarf galaxy candidates have been discovered in the vicinity of NGC\,253, that
   we named them Do\,V, Do\,VI, Do\,VII, Do\,VIII and Do\,IX. Assuming they are associated to NGC\,253, their total absolute $V$-magnitudes fall in the $-7$ to $-9$\,mag range, which is typical for dwarf satellites in the local Universe. The central surface brightness tend to be extremely low for all the discovered dwarfs and fall roughly in the range of 25--26\,mag\,arcsec$^{-2}$ in $g$-band.
   We present a new list of galaxy candidates located around the giant spiral NGC\,253.}
   {With the inclusion of these additional satellite candidates, the overall spatial distribution of the system becomes less flattened and is now broadly consistent with analogs drawn from $\Lambda$CDM expectations. Interestingly, the distribution appears to be rather lopsided. Yet, firm conclusions on the presence of absence of a correlated satellite structure are hampered since distance information is lacking, the census of observed dwarfs in the system remains far from complete, and spectroscopic velocities are not even available for most known satellites.}

   \keywords{galaxies: individual: NGC253 -- galaxies: formation -- galaxies:dwarf -- surveys}

\maketitle


\section{Introduction}

While the Lambda Cold Dark Matter ($\Lambda$CDM) model is extremely successful in describing the observable Universe on large scales (e.g., \citealt{PlanckCollaboration2020}), it has faced difficulties reproducing observations at the smaller scales, e.g.\ of nearby galaxy groups. 
Tensions remain between the number of observed satellite dwarf galaxies versus theoretical predictions.
To date, there are $\sim$63\footnote{According to the Local Volume Database, \url{https://github.com/apace7/local_volume_database},\citep{Pace2024arXiv241107424P}} known satellite galaxies of the Milky Way, yet simulations predict thousands of subhalos with large enough masses to form a satellite galaxy ($M_{\rm peak}\gtrsim10^7\,M_{\odot}$) \citep{2017ARA&A..55..343B}.
Recent studies give us some probable computational solutions 
\citep[see, for example, the dark-matter-only simulations and cumulative number counts of dwarf galaxies in the studies of][]{2014MNRAS.438.2578G,2017MNRAS.464.3108G}.
At the same time, considerable hope of the astronomical community still lies in the discovery of additional faint 
and ultra-faint satellites in the nearby galaxy groups. Here is a very incomplete list of works with recent discoveries: 
\citet{2017ApJ...837..136D, 2018MNRAS.474.3221M, 2023A&A...678A..16K, 2024Ap.....67..267K, 2024ApJ...966...89A}.

The other important problem is the existence of ``satellite planes'' preferably discovered in the nearby Universe.
The anisotropic distribution of Galactic satellites is well known~\citep{1976MNRAS.174..695L}
and is hardly explained within the modern theory of structure formation. 
Analysis of the system of satellites of the Milky Way led to the discovery of a planar structure of their distribution~\citep{2012MNRAS.423.1109P}.
The existence of the satellite plane contradicts the predictions of the subhalo distribution in the $\Lambda$CDM-theory~\citep{2018MPLA...3330004P}.
Later, similar planar structures were found around several nearby giant galaxies.
\citet{2013Natur.493...62I} found a thin plane in the Andromeda galaxy satellite system. 
\citet{2015ApJ...802L..25T} reported on the planar structure of the satellites around Centaurus\,A, later shown to also show a strong velocity coherence reminiscent of rotation \citealp{2018Sci...359..534M}. More recently, \citet{2024MNRAS.528.2805K} reported the discovery of a co-rotating satellite structure around NGC\,4490. 
The apparent prevalence of these planar structures challenges the $\Lambda$CDM paradigm \citep{2020MNRAS.491.3042P, 2021A&A...645L...5M}, but this conclusion is based on a limited number of hosts. Additionally, there are works that argue planar structures do not pose a challenge to the $\Lambda$CDM paradigm (see e.g. \citealt{Boylan-Kolchin2021}). Consequently, it is crucial to search for similar structures around additional hosts to understand the abundance, significance, and the degree of flatness of satellite planes. 
This aim requires overcoming the challenges of discovering additional dwarf galaxies, and measuring their distances with high precision.
Both tasks are fundamental for a better understanding of the galaxy distribution, 
cosmic structure formation, and the formation and evolution of galaxies in group and field environments.

One intriguing nearby structure that has so far been neglected in this context 
is a conglomeration of bright spiral galaxies in the constellation of Sculptor,
including NGC\,24, NGC\,45, NGC\,55, NGC\,247, NGC\,253, NGC\,300 and NGC\,7793.
Together with their dwarf satellites it forms a loose filament stretched along
the line of sight from the Local Group to a distance of about 7\,Mpc~\citep{1998AJ....116.2873J,2003A&A...404...93K}. 
The filament is located in the Local Supercluster plane almost in the Anti-Virgo direction. 
The central part of the filament is the group of dwarf galaxies around 
the luminous late-type spiral NGC\,253 at a distance of 3.70\,Mpc~\citep{2021AJ....162...80A}. 
This galaxy, also known as the Sculptor Galaxy, dominates the group.
Its luminosity, $\log L_K/L_{\odot} = 10.98$, 
exceeds the luminosity of the Milky Way or the Andromeda galaxy \citep{2021AJ....161..205K},
and it is 11 times brighter than the second brightest galaxy of the group - NGC\,247.
NGC\,253 is one of the nearest starburst galaxies 
with a present-day star formation rate of 5~$M_{\odot}$\,yr$^{-1}$ \citep{2002ApJ...574..709M}.
It is suggested that the starburst was triggered by a merger with a gas-rich dwarf in the past 200 Myr \citep{2010ApJ...725.1342D}.

\citet[][hereafter Paper~I]{2021A&A...652A..48M} claim the possible existence of flattened and velocity-correlated system of dwarfs around NGC\,253. 
This satellite plane is only 31~kpc thick with a minor-to-major axis ratio of 0.14.
If this is true, then it turns out that most of nearby giant galaxies, i.e. the Milky Way~\citep{2005A&A...431..517K}, Andromeda galaxy~\citep{2007MNRAS.374.1125M}, Centaurus\,A~\citep{2015ApJ...802L..25T}, M\,81~\citep{2013AJ....146..126C}, as well as NGC\,253~(Paper~I), have extended satellite planes, which would be extremely intriguing. 

At a distance of 3.7~Mpc, NGC\,253 is one of the closest spirals to the Local Group and thus the natural place to dig for LSB dwarf galaxies that could provide new insights on the presence satellites planes around nearby galaxies outside the LG. This member of the Sculptor group has been surveyed for satellite galaxies in the past \citep{1997AJ....114.1313C,1998A&AS..127..409K,1998AJ....116.2873J,2000A&AS..146..359K,2000AJ....119..593J,2014ApJ...793L...7S,2016ApJ...816L...5T, Mutlu-Pakdil2021, Carlsten2022, Mutlu-Pakdil2024, Okamoto2024}. 

In the last years, the exquisite imaging from the DESI Legacy Surveys~\citep{2019AJ....157..168D} have probed for the first time large-scale sky regions at very low surface brightness regime (28.0--29.0~mag\,arcsec$^{-2}$).  This has allowed the discovery of many very faint dwarf satellites around nearby galaxies using automatic detection algorithms~\citep{2021ApJS..252...18T} and systematic visual inspection of these public deep images~\citep{2021A&A...652A..48M, 2022AJ....163..234K, 2022AstBu..77..372K, 2023A&A...671A.141M}.
Three new dwarf galaxies, Do\,II, Do\,III and Do\,IV were reported in the vicinity of NGC\,253 in the framework of a systematic low surface brightness galaxy search. \cite{Mutlu-Pakdil2021} independently discovered DoII and presented its confirmation with HST data, while \citealt{Mutlu-Pakdil2024} provided confirmation for the candidates discussed in Paper 1 as well as for the other two satellites, which previously lacked distance information. Their total absolute magnitudes, transformed to the $V$-band, fall in the range from about $-7$ to about $-9$~mag at the distance of NGC\,253, which are typical for satellite dwarf galaxies in the Local Universe. 

In this paper, we present the discovery of five low-surface brightness dwarf galaxies around the NGC\,253--NGC\,247 group in the Sculptor group using DESI Legacy Surveys imaging data. With this updated census of low-mass systems, we revise our previous result on the possible existence of a spatially flattened and velocity-correlated dwarf galaxy system around NGC\,253.
  
\begin{figure*}
   \begin{center}
        \includegraphics[width=\textwidth]{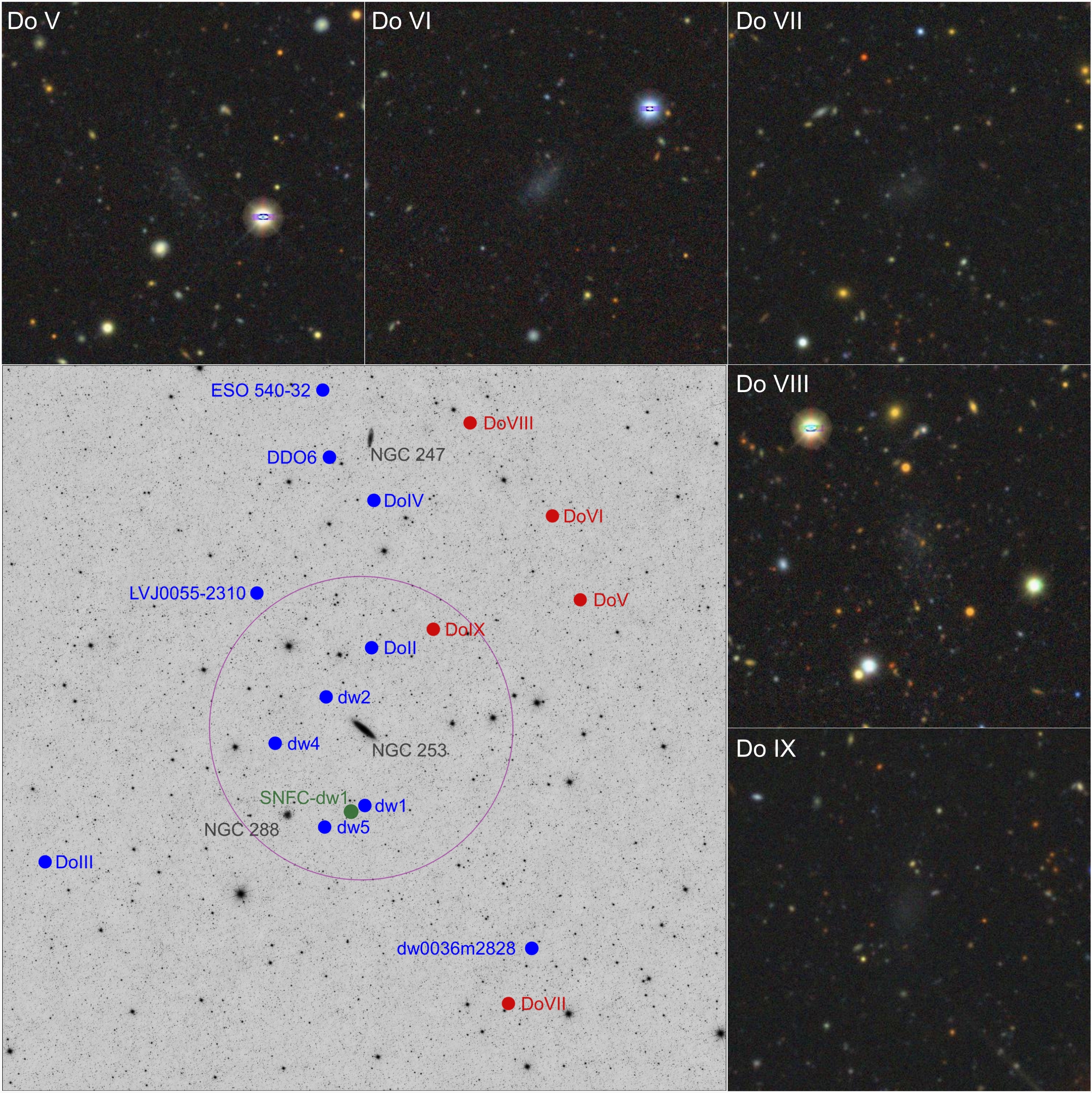}
    \caption{ {\it Left panel}: Position of the six dwarf galaxy candidates (solid red circles) reported in this study with respect to the spiral NGC\,253. The purple circular line corresponds to the area explored by the PISCes survey~\citep{2016ApJ...816L...5T} extending up to $\sim$ 150 kpc from the centre of NGC\,253. The total field of view of this image is 15$\fdg$ $\times$ 15$\fdg$. {\it Top and right panels}: Full colour version of the image cutouts obtained with {\it legacypipe} for Do\,V, VI, VII, VIII and IX. North is up and east is left. The field of view of all these image cutouts is $2.5\arcmin \times 2.5\arcmin$.}
    \label{fig-map-ngc253}
     \end{center}
\end{figure*}   

\section{SEARCHING STRATEGY AND DATA ANALYSIS}

\subsection{Searching strategy and Image cutout data }

The dwarf galaxy satellite candidates reported in this paper were found by the amateur astronomer Giuseppe Donatiello (abbreviated as Do) by visual inspection of the Dark Energy Camera~\citep[DECam,][]{2015AJ....150..150F} images of the Scuptor group of galaxies  available from the DESI Legacy Imaging Surveys~\citep{2019AJ....157..168D}. 
A total number of 13 candidates were detected in an total explored area of $15\times 10$ degrees. These systems were not identified in previous automatic searches of diffuse stellar systems based on resolved stellar source density maps extracted from these data~\citep[e.g.,][]{2021ApJS..252...18T}. The typical angular resolution of these data (estimated from the seeing of the images) is $\sim$ 0.9\arcsec. 

For this paper, we only focus on those brighter candidates which DESI Legacy Surveys images were suitable for a feasible photometry and structural analysis (see the next Subsection), following the same approach described in Paper~I. The given names with their corresponding positions are given in Table~\ref{tab:dwarf_coordinates}.

\begin{table}
    \caption{Coordinates of the five newly detected satellite candidates.}
    \label{tab:dwarf_coordinates}
    \centering
    \begin{tabular}{lccr}
    \hline \hline
    Name & \multicolumn{2}{c}{RA\hfil (J2000)\hfil DEC} &  Discovered on\\
         & [h m s]           & [$^\circ$ $^\prime$ $^{\prime\prime}$]     & [M-D-Y] \\
    \hline
    Do V   & $00\,32\,58.32$ & $-23\,16\,45.1$ & 09--06--2021\\
    Do VI  & $00\,34\,55.68$ & $-21\,59\,41.3$ & 05--20--2022\\
    Do VII & $00\,37\,34.82$ & $-29\,28\,44.8$ & 10--19--2022\\
    Do VIII& $00\,40\,34.90$ & $-20\,33\,25.2$ & 11--20--2020\\
    Do IX  & $00\,42\,42.24$ & $-23\,46\,10.6$ & 07--27--2020\\
    \hline
    \end{tabular}
    
\end{table}

Image cutouts centered on each satellite candidate were  obtained by coadding DESI Legacy Surveys images of these systems  using the \textsc{legacypipe} software of the DESI Legacy imaging surveys~\citep[see, for example, Fig.~2 in ][]{2023A&A...671A.141M}. A color version of the resulting coadded image cutouts of these five dwarf galaxies are shown in Fig.~\ref{fig-map-ngc253} (top and right panels). The surface brightness limit of these images in the vecinity of NGC\,253 region are $\mu=29.3$, 29.0 and 27.7~mag\,arcsec$^{-2}$ for the $g$, $r$ and $z$ bands respectively measured as 3$\sigma$ in 10$\times$10~arcsec boxes, as estimated in Paper~I. 


\begin{figure*}
        \includegraphics[width=0.9\textwidth]{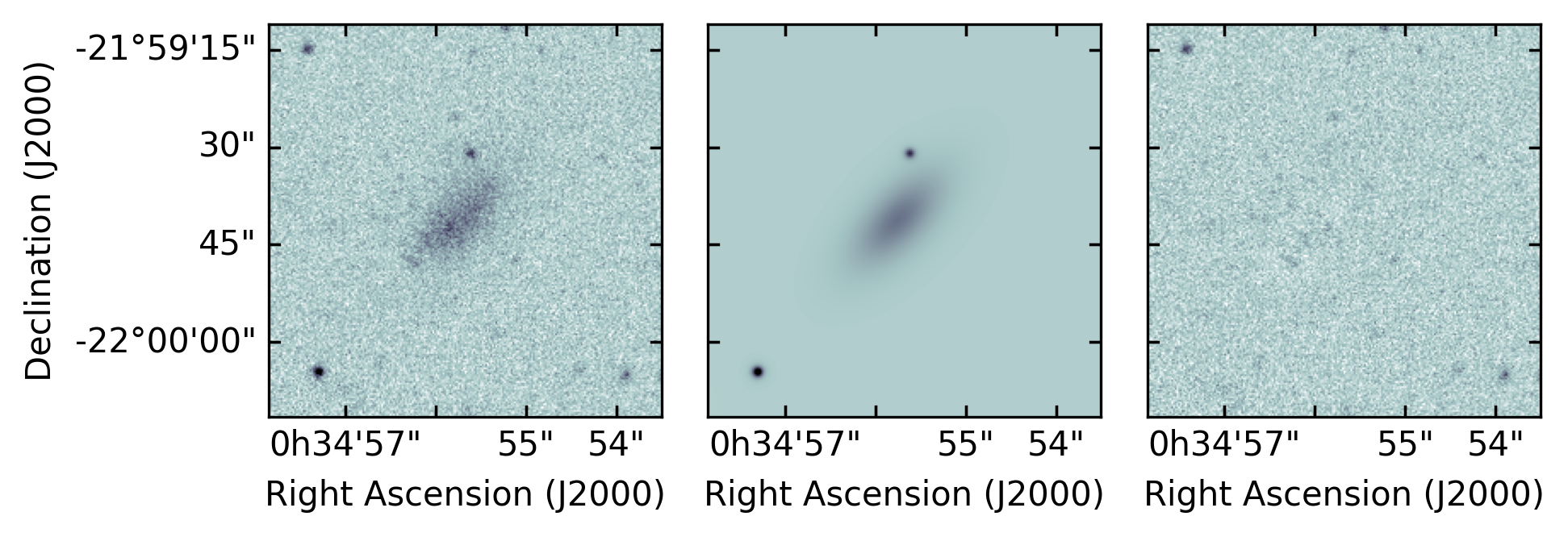}
        \includegraphics[width=0.9\textwidth]{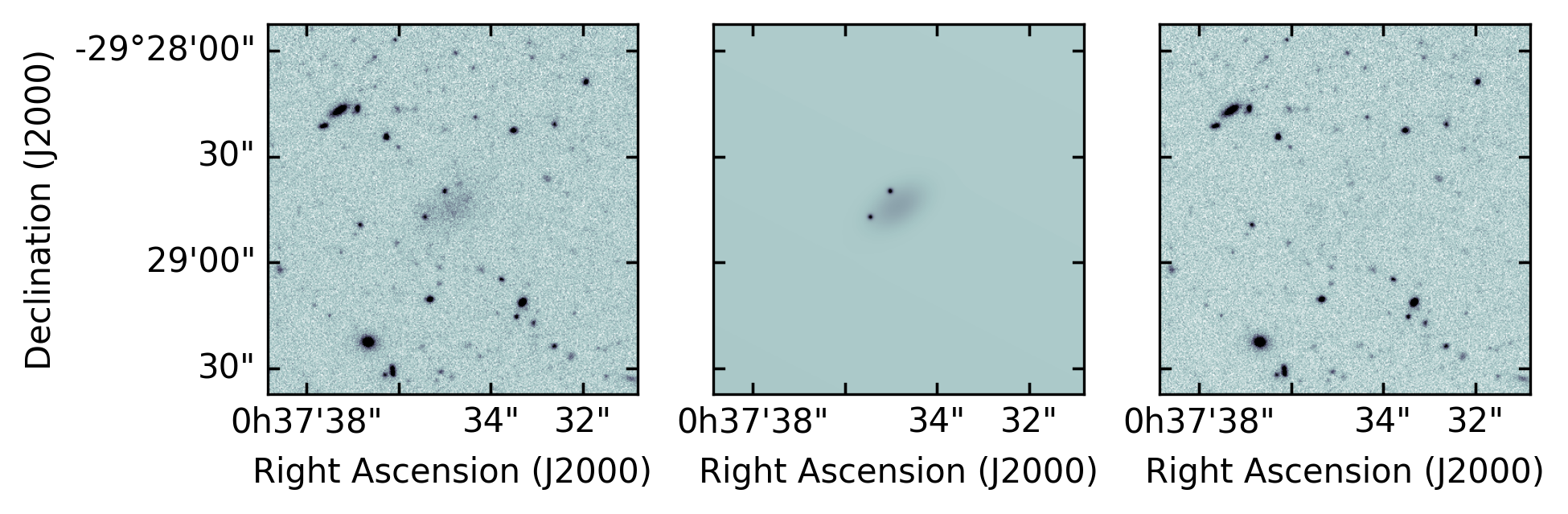}

    \caption{Light profile GALFIT modeling results \textit{Top Row:} $g$-band for Do\,VI; \textit{Bottom Row:} $r$-band for Do\,VII.}
   \label{fig:resolution}
\end{figure*}

\begin{table*}
    \caption{Photometric and structural properties for the five satellite candidates derived by fitting one component Sersic profiles to the $g$- and $r-band$ images. Braces indicate that a parameter was fixed during the fitting process.}
    \label{tab:galfit_results}
    \centering
    \begin{tabular}{lrrrrrr}
\hline\hline
Name& Do V & Do VI & Do VII & Do VIII & Do IX \\
\hline
$m_g\,\mathrm{[mag]}$ & $20.39 \pm 0.04$ & $19.71 \pm 0.01$ & $20.68 \pm 0.02$ & $19.24 \pm 0.08$ & $20.02 \pm 0.04$ \\
$\mu_{g, \mathrm{eff}}\,\mathrm{[mag\,arcsec^{-2}]}$ & $27.00 \pm 0.03$ & $25.78 \pm 0.01$ & $26.68 \pm 0.05$ & $27.60 \pm 0.02$ & $27.62 \pm 0.02$ \\
$n_g$ & $0.84 \pm 0.07$ & $0.64 \pm 0.02$ & $0.43 \pm 0.03$ & $1.09 \pm 0.08$ & $0.76 \pm 0.05$ \\
$\mathrm{PA}_g$\,[deg] & $28 \pm 2$ & $-44.0 \pm 0.7$ & $-57 \pm 2$ & $12.9 \pm 0.8$ & $-27 \pm 2$ \\
$(b/a)_g$ & $0.50 \pm 0.02$ & $0.469 \pm 0.005$ & $0.53 \pm 0.02$ & $0.311 \pm 0.009$ & $0.43 \pm 0.01$ \\
$R_{g, \mathrm{eff}}$\,[arcsec]& [9.0] & $7.6 \pm 0.1$ & $7.6 \pm 0.3$ & $24 \pm 2$ & $15.5 \pm 0.8$ \\
$m_r\,\mathrm{[mag]}$ & $20.28 \pm 0.04$ & $19.37 \pm 0.02$ & $20.32 \pm 0.03$ & $18.85 \pm 0.07$ & $19.60 \pm 0.04$ \\
$\mu_{r, \mathrm{eff}}\,\mathrm{[mag\,arcsec^{-2}]}$ & $26.31 \pm 0.03$ & $25.41 \pm 0.01$ & $26.08 \pm 0.03$ & $26.99 \pm 0.02$ & $27.02 \pm 0.03$ \\
$n_r$ & $0.70 \pm 0.07$ & $0.64 \pm 0.02$ & $0.45 \pm 0.04$ & $1.22 \pm 0.07$ & $0.78 \pm 0.06$ \\
$\mathrm{PA}_r$\,[deg]& $34 \pm 1$ & $-42.1 \pm 0.9$ & $-58 \pm 3$ & $10 \pm 1$ & $-18 \pm 1$ \\
$(b/a)_r$& $0.31 \pm 0.01$ & $0.486 \pm 0.006$ & $0.59 \pm 0.02$ & $0.41 \pm 0.01$ & $0.39 \pm 0.01$ \\
$R_{r, \mathrm{eff}}$\,[arcsec] & $9.0 \pm 0.5$ & $7.3 \pm 0.1$ & $6.5 \pm 0.3$ & $18 \pm 1$ & $15.5 \pm 0.8$ \\
\hline
\end{tabular}
\end{table*}

\subsection{Photometric Analysis}
After creating the image cutouts, the images are further processed. Here, we make use of the \texttt{GNU Astronomy Utilities} \citep[][\texttt{gnuastro}, \url{https://ascl.net/1801.009}]{gnuastro}. During images processing, the following utilities are incorporated: \texttt{astarithmetic}, \texttt{astcrop}, \texttt{aststatistics}, \texttt{asttable}, \texttt{astscript-psf-select-stars}, \texttt{astscript-psf-stamp}, \texttt{astscript-psf-scale-factor}, \texttt{astscript-psf-unite} \citep{gnuastro,noisechisel_segment_2019,gnuastro-psf}. First, we find the saturation level of the image and mask out all saturated pixels. This step is necessary for building up the empirical model of the extended PSF (ePSF).


\subsubsection{ Photometric and Structural Properties}

To build up the ePSF models ($g$- and $r$-band), first we define a two distinct sets of stars: Bright stars (stars that show saturation in center, $m_{\mathrm{star}}<15.5$) and faint stars ($16 < m_{\mathrm{star}}<20$). The faint stars are stacked to build the inner part of the ePSFs, while the set of bright stars builds up the outer regions\footnote{The exact procedure is explained in the \texttt{gnuastro} manual: \url{https://www.gnu.org/software/gnuastro/manual/html_node/Building-the-extended-PSF.html}.}. In total, we use 498 faint stars for the inner part of the ePSFs and 56 bright stars for the outer part.

To extract reliable structural parameters for our satellite galaxy candidates, we fit the two dimensional light distribution of all candidates with one component Sersic profiles, by using \texttt{galfit} \citep{galfit1, galfit2}. Here, we use our ePSF kernels for the convolution of the model images. If there are bright sources in the image, close to the satellite candidate, we also fit these to reduce their impact on our low surface brightness targets. We start the fitting process in $r$-band and use the best fitting parameters as initial parameters for the $g$-band fitting. 
To derive the effective surface brightness of our targets, we rerun \texttt{galfit}, using the \texttt{sersic2} function (this function is similar to the standard \texttt{sersic} model, but returns the effective surface brightness instead of an integrated magnitude per source), while fixing all other parameters to the best fitting values of the previous run. For Do\,V, we had to fix the effective radius that was found in $r$-band during the fitting of the $g$-band data, to find a solution. In Table~\ref{tab:galfit_results}, we list the results of the the fitting process for all galaxies in $g$- and $r$-band\footnote{Parameter errors were estimated by \texttt{galfit}}.

In order to compare the detected dwarf galaxy candidates with already confirmed local group dwarf galaxies, we calculate $V$-band magnitudes and surface brightness values based on the transformation given by \citet{desdr2}.

\begin{equation}
V =
  \begin{cases}
    g - 0.465(g - r) - 0.020,  &  \mbox{if }     -0.5 < (g - r) \leq 0.2 \\
    g - 0.496(g - r) - 0.015,  &  \mbox{if }\quad 0.2 < (g - r) \leq 0.7
  \end{cases}
\end{equation}

\begin{figure}
    \centering
    \begin{subfigure}{0.9\linewidth}
       \includegraphics[width=\hsize]{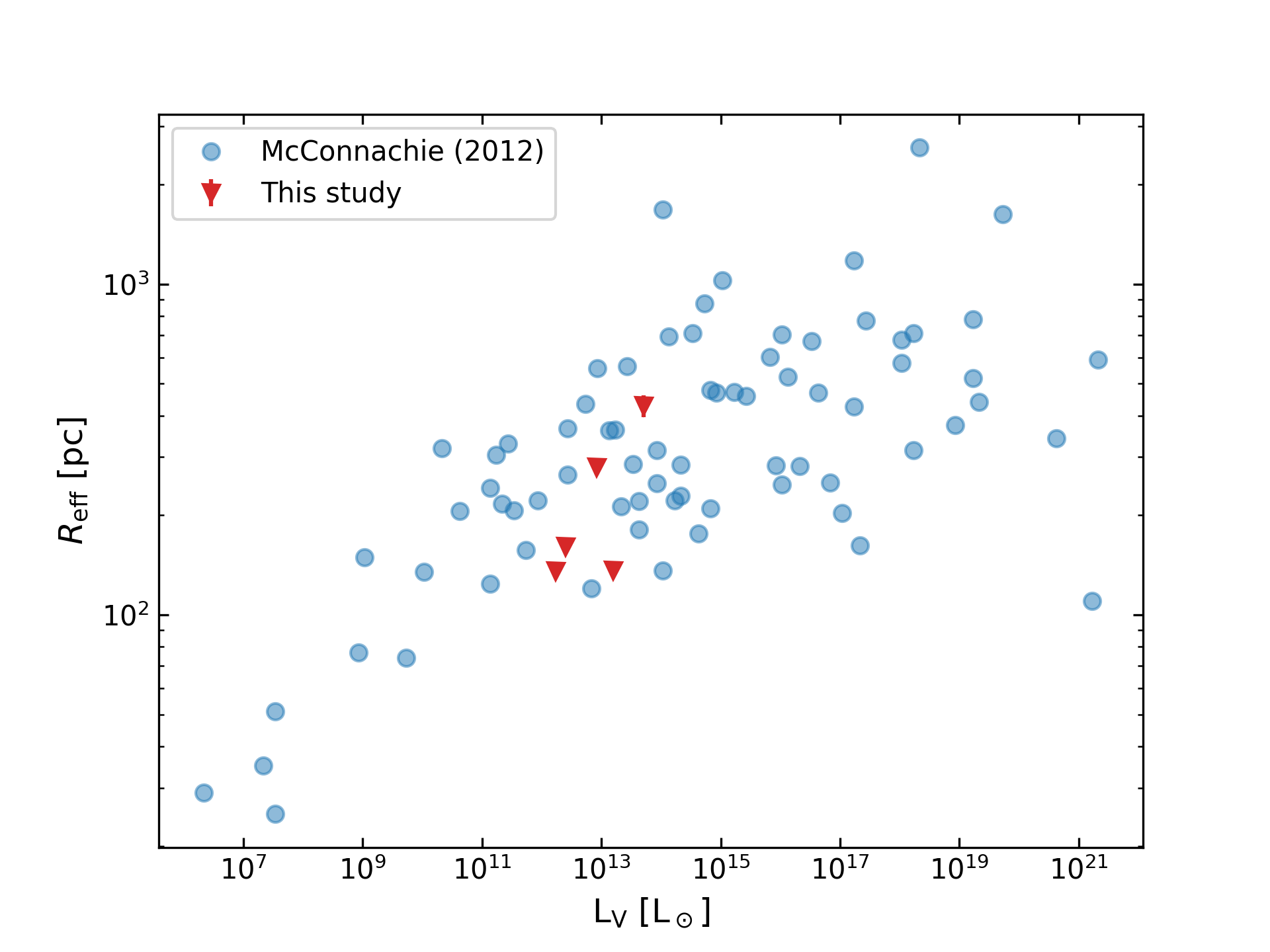}
    \end{subfigure}
    \\
   \begin{subfigure}{0.9\linewidth}
       \includegraphics[width=\hsize]{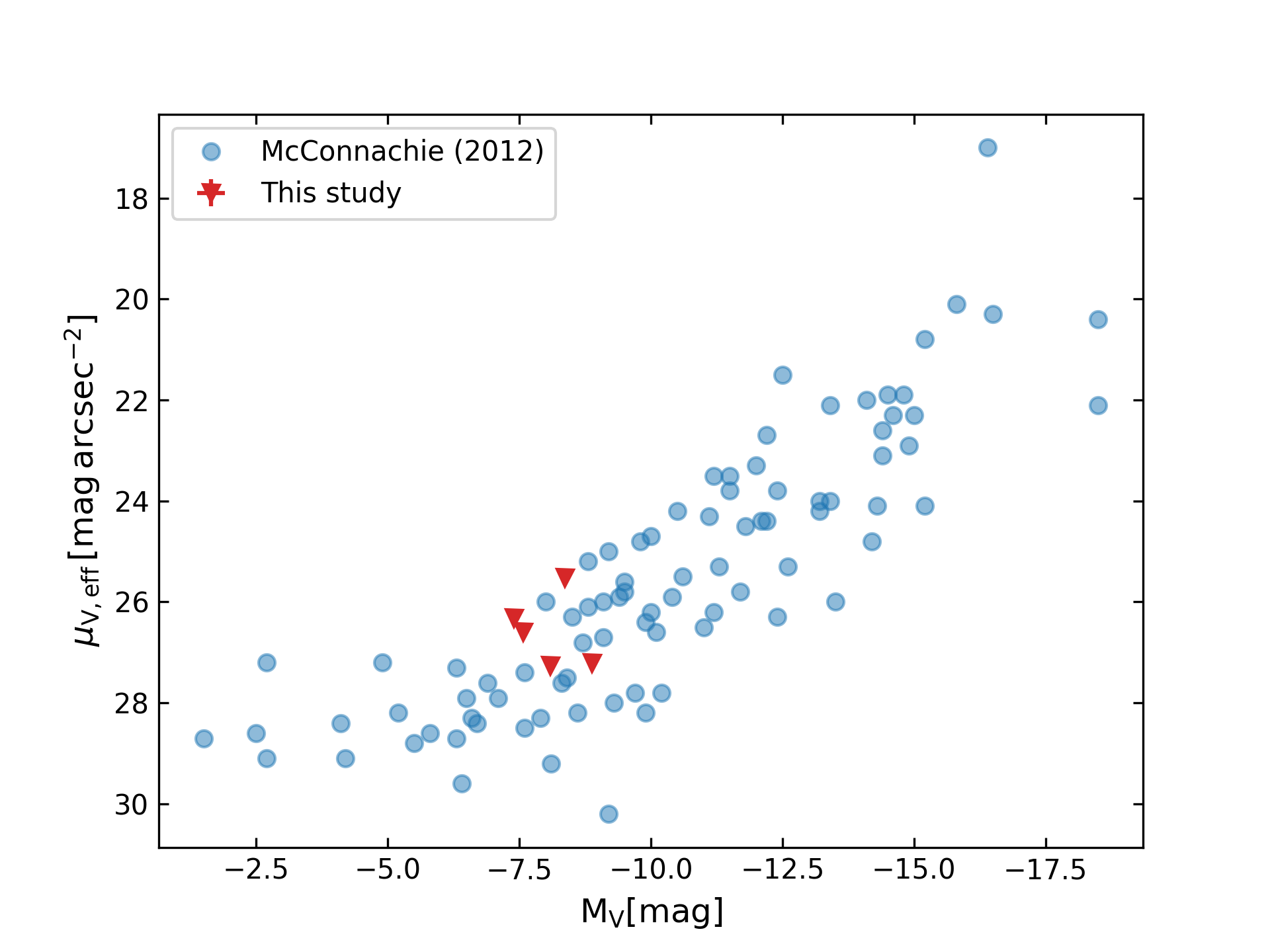}
    \end{subfigure}
    \\
    \begin{subfigure}{0.9\linewidth}
       \includegraphics[width=\hsize]{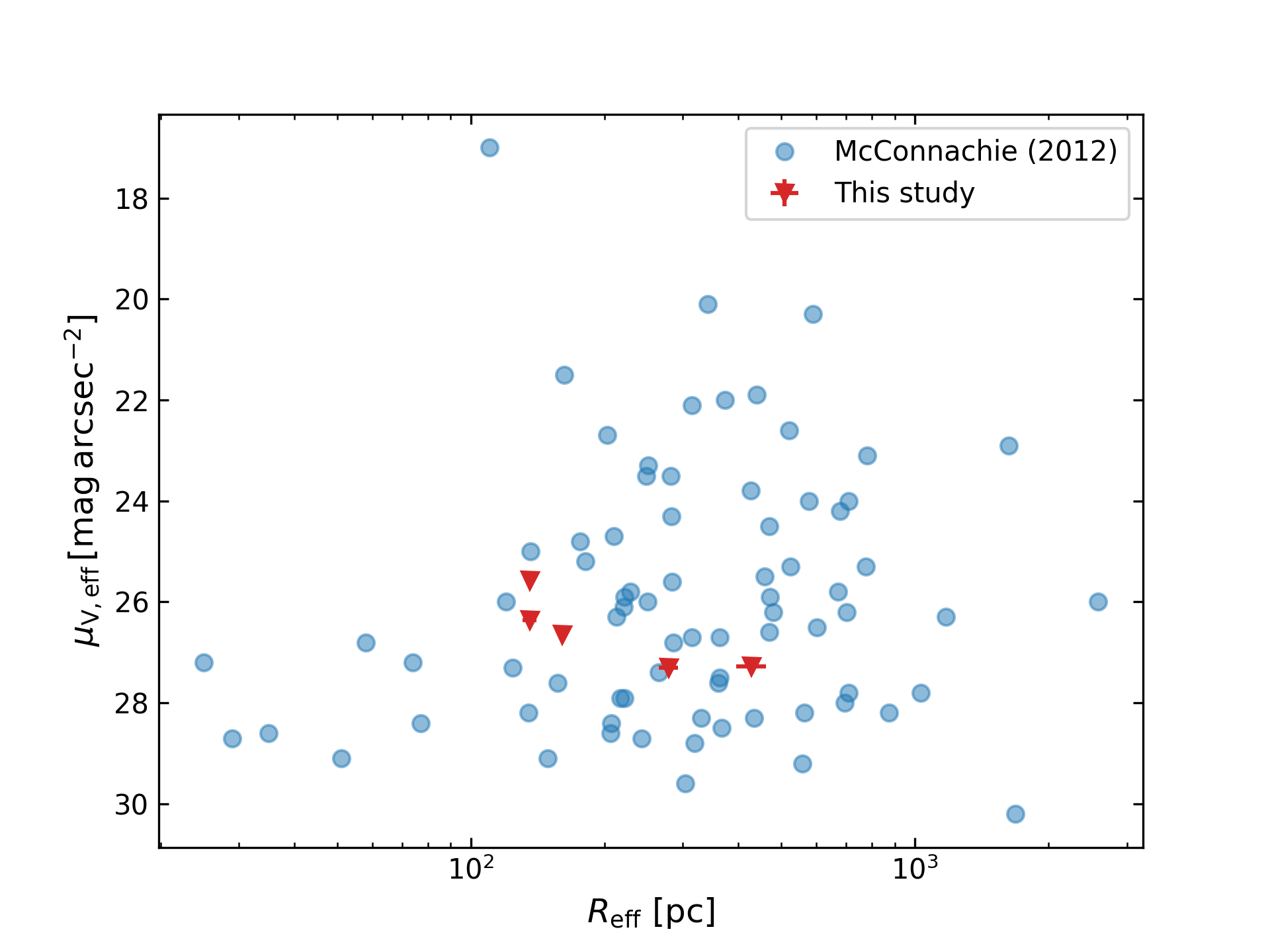}
    \end{subfigure}
     \\
    \caption{Comparison of the satellite dwarf galaxy candidates against confirmed Local Group dwarf galaxies.}
    \label{fig:local_group dwarf_gal_comp}
\end{figure}

In Table \ref{tab:Vmag}, we present the derived surface brightness values as well absolute magnitudes and the physical extent of each dwarf galaxy candidates, assuming that they have the same distance as NGC~253. In the three panels in Fig. \ref{fig:local_group dwarf_gal_comp}, we compare the derived photometric properties with the local group dwarf galaxy sample described in \citep{McConnachie2012}. Under the assumption that our dwarf galaxy candidates are part of the NGC~253 group, we find good agreement between the photometric properties of our targets and the local group dwarf galaxies.

\begin{table}
    \caption{Effective $V$-surface brightness levels (converted from $g$- and $r$-band measurements), absolute $V$-band magnitudes, $g$-band effective radii converted to parsec , and galactic extinction corrections based on \citet{gal_ext} for all observed satellite candidates.}
    \label{tab:Vmag}
    \centering
    \begin{tabular}{lrrrr}
\hline\hline
Target & $\mu_{\mathrm{V, eff}}$ & $\mathrm{M_V}$ & 
\multicolumn{1}{c}{$\mathrm{R_{g, eff}}$} & $\mathrm{A_V}$ \\
        & [mag\,arcsec\textsuperscript{-2}] & [mag] &
\multicolumn{1}{c}{[pc]} & [mag]\\    
\hline
Do V    & $26.60 \pm 0.03$ & $-7.57 \pm 0.03$ & $161 \pm 9$ & 0.05 \\
Do VI   & $25.52 \pm 0.01$ & $-8.37 \pm 0.02$ & $136 \pm 2$ & 0.06 \\
Do VII  & $26.32 \pm 0.03$ & $-7.40 \pm 0.02$ & $136 \pm 5$ & 0.05 \\
Do VIII & $27.21 \pm 0.02$ & $-8.88 \pm 0.05$ & $429 \pm 32$ & 0.07 \\
Do IX   & $27.27 \pm 0.02$ & $-8.09 \pm 0.03$ & $279 \pm 14$ & 0.04 \\
\hline
\end{tabular}
\end{table}

\section{DISCUSSION}

\subsection{Updating the Sculptor group dwarf galaxy census}

In Paper~I we gave a detailed description of the structure of the NGC\,253 group. We estimated the NGC\,253 group mass of about $8\times10^{11}$~M$_\sun$, virial and so-called turn-around radii, equal to $R_{200}=186$ and $R_\mathrm{ta}=706$~kpc respectively.
Since then, there have been no changes in known velocities, but TRGB distances have been measured for a large number of dwarfs and the group has been replenished with new members and candidates.
We summarize the current status of the group in Table~\ref{tab:ngc253group}.
It lists galaxies less than $15\fdg5$ away from NGC\,253, which corresponds to a projected distance of 1~Mpc.
The galaxies are ordered according to their projected distance.
Within 500~kpc, NGC\,253 contains 14 confirmed satellites and 5 new candidates.
A group of galaxies around NGC\,7793 at $3.63$~Mpc~\citep{2021AJ....162...80A} is located at the boundary of the system and just beginning their infall to the NGC\,253 group.
According to TRGB-distance measurements, the galaxies DDO\,226 and NGC\,59, although projected onto the NGC\,253 group, are located outside its turn-around radius.
Note that Sculptor\,SR turned out to be a background galaxy at a distance of about 19~Mpc~\citep{Mutlu-Pakdil2024}.

\begin{figure*}
\centering
\includegraphics[width=\textwidth]{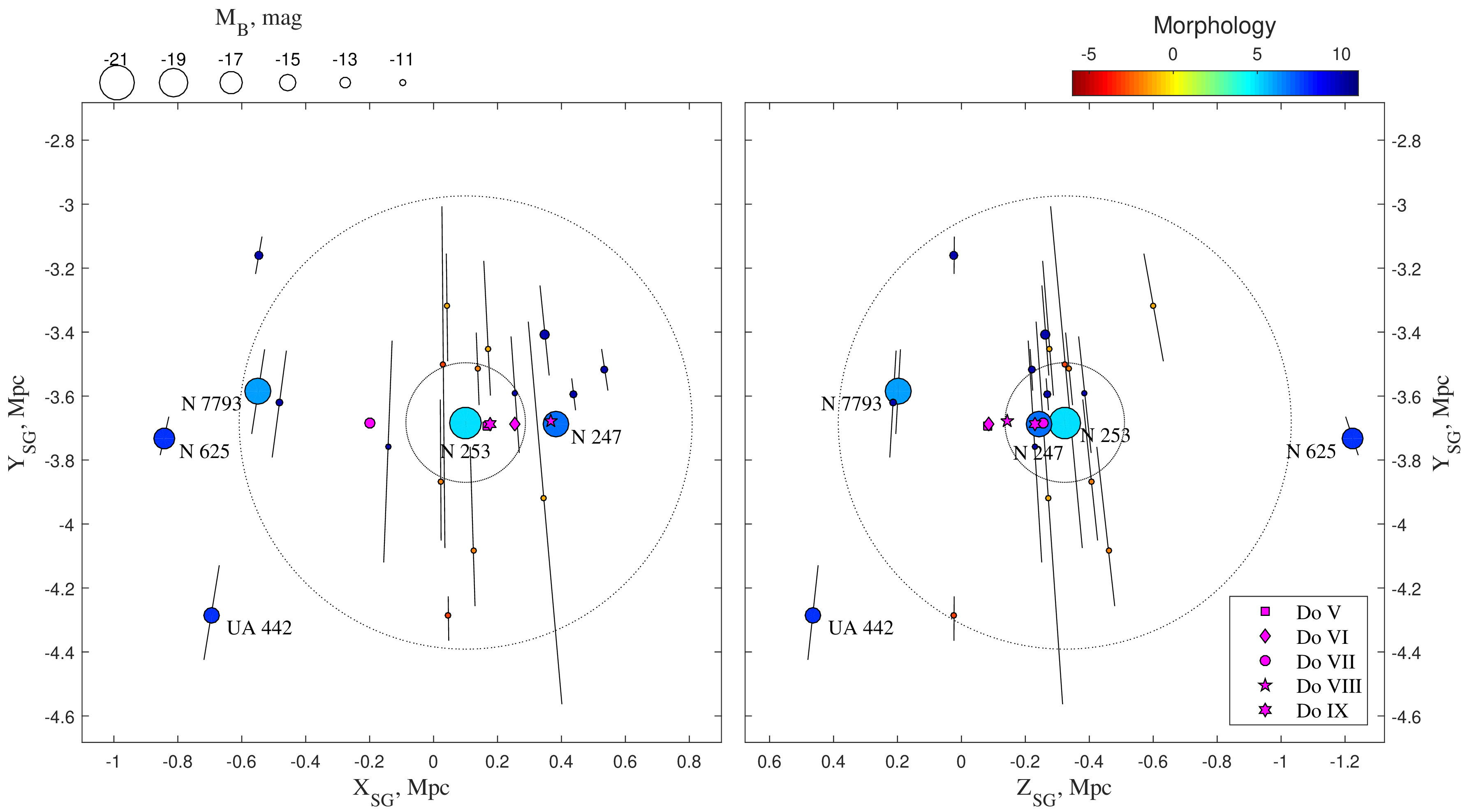}
\caption{
Distribution of galaxies around NGC 253 in the supergalactic coordinates. 
\textit{Left-hand panel} shows the projection on the plane of the Local Supercluster, 
while \textit{right-hand panel} presents its edge-on view.
The sizes and colors of the dots reflect the $B$-band absolute magnitude and morphology of the galaxies, according to the legends above.
The line segments correspond to the distance errors.
The dotted circles mark the virial zone of $R_{200}=190$ and the turn-around radius of $R_\mathrm{ta}=710$~kpc around NGC\,253.
The expected positions of the discovered galaxies are indicated by magenta symbols.
}
\label{fig:ngc253xyz}
\end{figure*}

Increasing the number of galaxies with known distances from 9~(Paper~I) to 15~(this compilation) allows us to build a more accurate map of the group (Fig.~\ref{fig:ngc253xyz}).
Note that the typical accuracy of distance estimates is comparable to the size of the virial zone around the central galaxy.
Nevertheless, it is clear that the distribution of satellites in the plane of the Local Supercluster and perpendicular to it has different widths.
The map illustrates the feature noted in Paper~I that the satellites form a flat structure lying close to the plane of the Local Supercluster.
The TRGB-distance of dw0036$-$2828, measured by \citet{Mutlu-Pakdil2024}, confirms its membership in the NGC\,253 group and significantly increase the satellite plane extent.
Obviously, Do\,VII and IX can belong to the same plane, while Do\,V, VI and VIII should lead to its thickening.

\subsection{Revising the existence of a plane of Satellite Galaxies around NGC~253}

\begin{figure}
\centering
\includegraphics[width=0.49\textwidth]{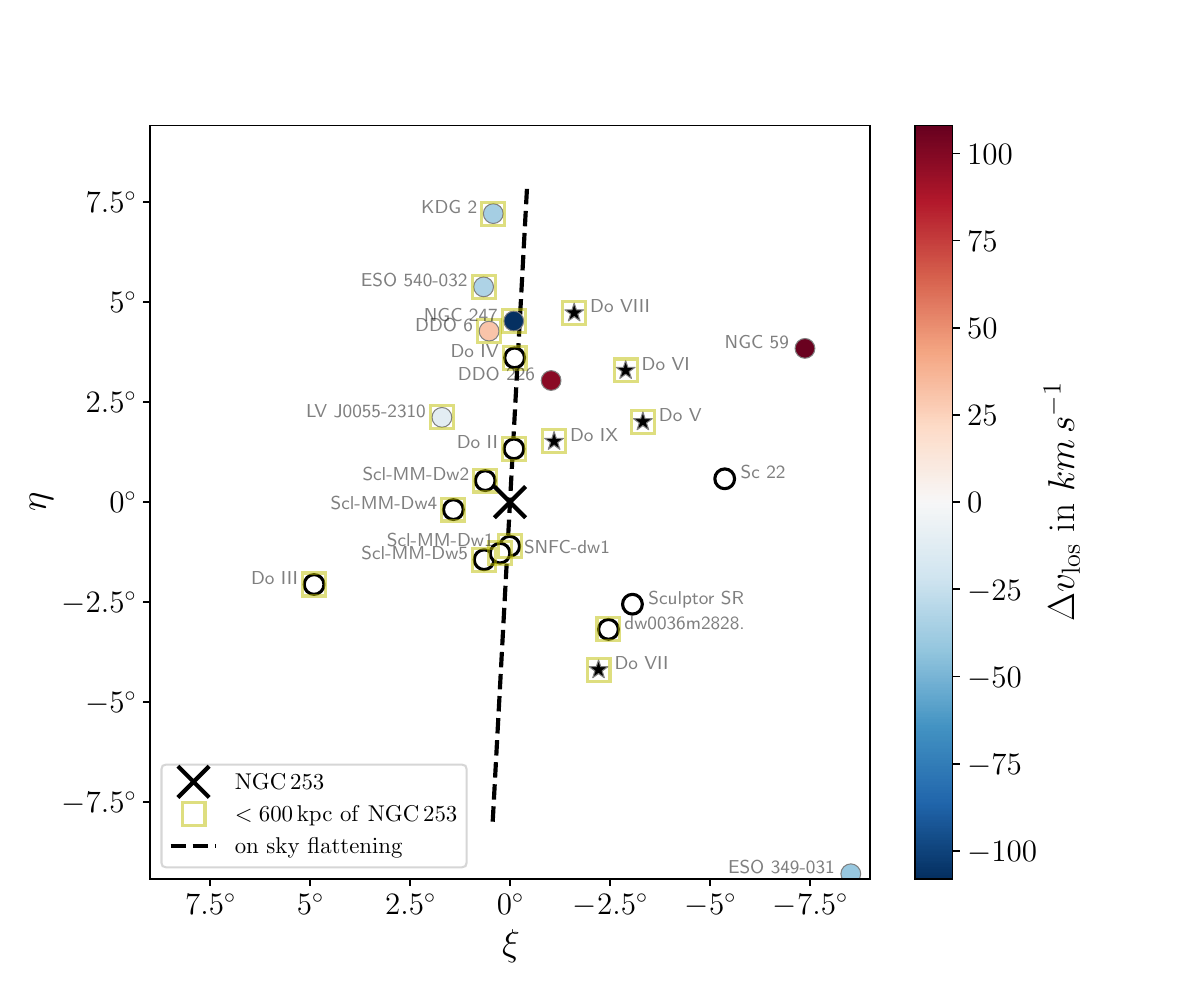}
\caption{
On-sky distribution of the galaxies around NGC\,253 in equatorial coordinates relative to the position of NGC\,253 (black cross). 
Galaxies with available velocity measurements are color-coded by their line-of-sight velocity component relative to that of NGC\,253. Galaxies with available distance measurements are plotted as open white circles, while our new candidates are plotted as filled black stars. Yellow boxes mark objects within 600\,kpc of NGC\,253, either measured in three dimensions if distances have been measured, or in projection for those objects for which such data is not available. The on-sky orientation of the major axis of the spatial distribution of these highlighted objects is shown as a dashed black line.
}
\label{fig:onsky}
\end{figure}

\begin{figure*}
    \centering
    \includegraphics[width=0.24\textwidth]{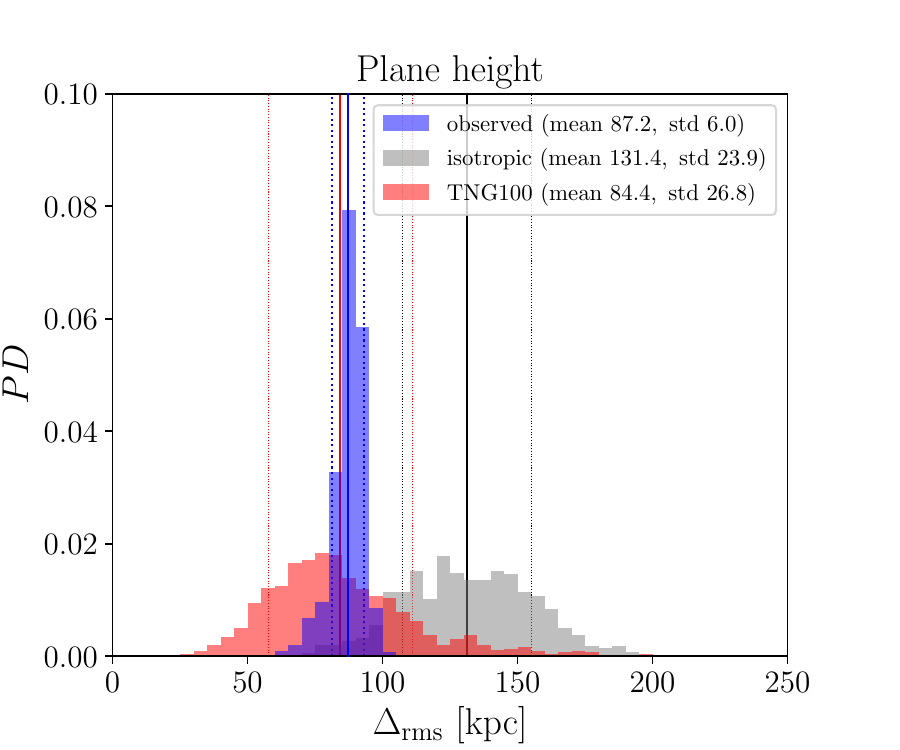}
    \includegraphics[width=0.24\textwidth]{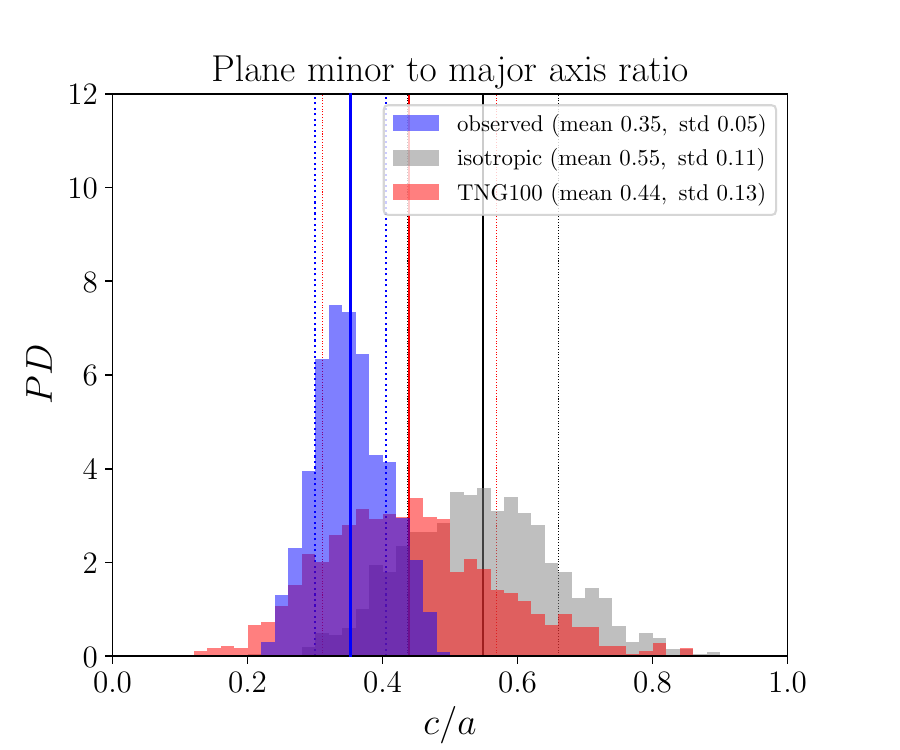}
    \includegraphics[width=0.24\textwidth]{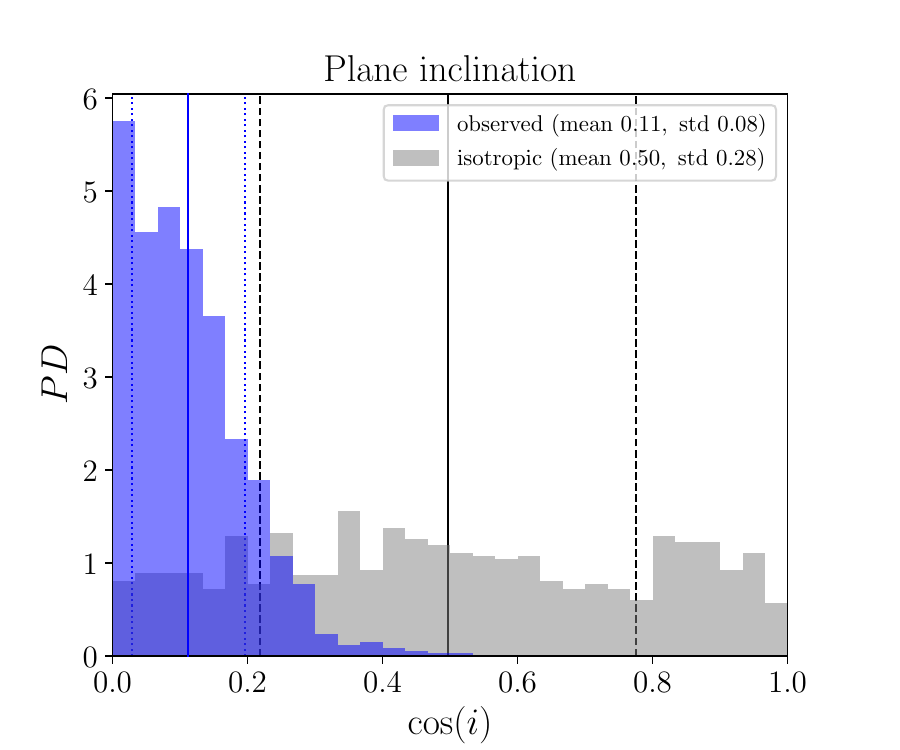}
    \includegraphics[width=0.24\textwidth]{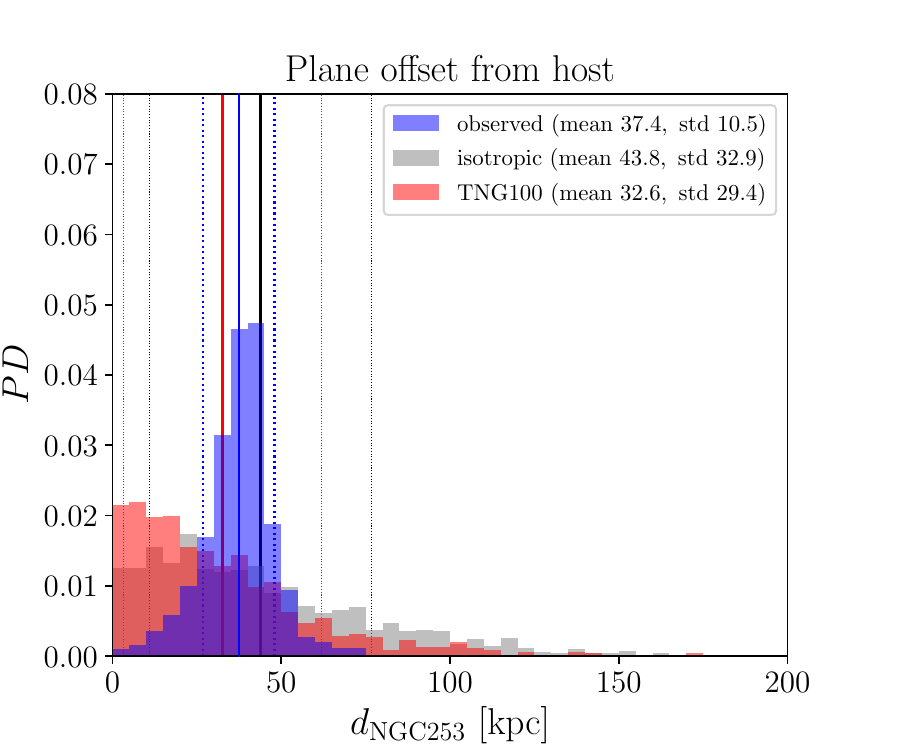}
    \caption{Parameters of  the ToI fits to the spatial distributions of galaxies with measured distances, that reside within 600\,kpc of NGC\,253. Shown are $\Delta_\mathrm{rms}$, the absolute rms plane height (far left); $c/a$, the minor-to-major axis ratio (middle left); $i$, the inclination of the best-fit plane with the line-of-sight (middle right); and $d_\mathrm{NGC\,253}$, the perpendicular offset of the best-fit plane from NGC\,253 (far right).  Blue shows the realizations drawing from the observed galaxy distances. They are more flattened than randomized samples drawn from isotropic distributions with identical radial distributions (shown in grey). However, they appear to be consistent with the parameters of satellite systems in the cosmological IllustrisTNG-100 simulation (shown in red). The mean and standard deviations of the shown distributions are given in each panel.}
    \label{fig:planefit}
\end{figure*}

The increase of the number of galaxies around NGC\,253 for which distances are available allows us to revisit the issue of a potential plane of satellites in this system. To do so, we update our comparison to systems selected from cosmological simulations, following same procedure as outlined in Paper~I. However, we now consider the 14 satellites with measured distances, up from only seven before (see Fig.~\ref{fig:onsky}). We again study the three-dimensional distribution via a tensor-of-inertia (ToI) fitting method, accounting for distance uncertainties with 1000 Monte-Carlo realizations. Figure~\ref{fig:planefit} shows the results of our plane fits. 

As before, we consider all dwarfs within 600\,kpc, thereby including only those 14 with measured distances. Overall, the relative and absolute flattening of the distribution becomes much wider. With the new sample of objects, we find a root-mean-square (rms) height from the best-fit plane of $\Delta_\mathrm{rms} = 87 \pm 6$~kpc, and a minor-to-major axis ratio of $c/a = 0.35 \pm 0.05$. These values have almost tripled from the ones based on the earlier, smaller sample in Paper~I ($\Delta_\mathrm{rms} = 31 \pm 5$~kpc and $c/a = 0.14 \pm 0.03$). 

The normal direction to the best-fit plane, however, remains in a similar orientation as before and is still aligned well with the Supergalactic plane. It also remains within $(19 \pm 6)^\circ$\ to the $\vec{e}_1$-direction, where $\vec{e}_1$\ is the eigenvector of the tidal tensor at the position of NGC\,253 which corresponds to the axis along which material in the cosmic web is compressed fastest. Such alignment has been found for several nearby satellite structures~\citep{2018MNRAS.473.1195L}. Similarly, the best-fit plane remains close to an edge-on orientation with an inclination of $i = (84 \pm 5)^\circ$.

Completely random systems drawn from isotropy, but reproducing the observed system's radial distribution, do typically result in wider distributions (gray in Fig.~\ref{fig:planefit}), with $\Delta_\mathrm{rms} = 131 \pm 24$\,kpc, and $c/a = 0.55 \pm 0.11$. However, given the wide spread in these parameters, the randomized and observed distributions overlap at the 1--2$\sigma$-level.

As in Paper~I, we also compare to satellite systems selected from the cosmological hydrodynamical simulation IllustrisTNG-100 \citep{2018MNRAS.475..676S, 2019ComAC...6....2N}. We refer to that paper for the details, noting only that we now select the 14 top-ranked galaxies per host to ensure the same number as in the observed sample.
As analogs, we again select host galaxies by virial mass, confined in the interval from 0.6 to $1.0 \times 10^{12}\,M_\sun$. We reject those with another galaxy of virial mass $>0.5 \times 10^{12}\,M_\sun$ within 1.2\,Mpc to ensure sufficient isolation.  All galaxies with a separation of 600\,kpc or less around the hosts are ranked, first by brightness and then by mass for those not containing stars. The flattening of the 13 top-ranked galaxies surrounding the host is measured. The resulting distributions in flattening and host offset are included in red in Fig.~\ref{fig:planefit}.

The $\Lambda$CDM satellite systems are more flattened than the randomized systems, but their distribution comfortably encompasses the flattening of the observed NGC\,253 system. This is the case both in absolute flattening ($\Delta_\mathrm{rms} = 84 \pm 27$\,kpc), and in relative axis ratio ($c/a = 0.44 \pm 0.13$). If we repeat the analysis considering only objects within 300\,kpc of the host, these results do not change qualitatively. Also within the smaller volume, there is no strong evidence for a substantially flattened overall distribution of the satellite system. Also the offset of the best-fit planes are broadly consistent among the observed, randomized, and simulated systems.


While the conclusion of \citet{Mutlu-Pakdil2024} was not based on a quantitative comparison, we have now measured the overall spatial flattening and compared it with satellite systems selected from a full cosmological context. We find that the overall flattening of the system is consistent with these model expectations, and thus we can now quantitatively confirm that there is not strong evidence for a spatially flattened satellite plane in this system anymore.

We note that these results are for the total dwarf system, but do not account for possible subsamples. For example, the significant M\,31 satellite plane is only composed of about half of the satellites, and proper motions indicate that a similar fraction can be expected to be actual satellite plane members for the Milky Way \citep{2021ApJ...916....8L, 2024A&A...681A..73T}. However, as of now the numbers of satellites is too low and insufficient number of spectroscopic velocities are available to attempt a meaningful subsample-analysis.

Maybe even more important is that the three more famous cases of satellite planes, those of the Milky Way, M\,31, and Centaurus\,A, caution us to not over-interpret this result. This is because in all three cases, the main evidence for cosmologically rare structures comes from the observed coherence of velocities. Such data are not available for most objects here, and of the five objects with available spectroscopic velocities, four do show a coherent velocity trend. Thus, since the spatial flattening remains close to edge-on (even if it is not as prominent as before), spectroscopic velocities remain a very promising avenue to shed light on this system and its status in regards to the issue of planes of satellite galaxies.

Beyond satellite planes, other phase-space correlations in systems of satellite galaxies are being increasingly investigated \citep[see][for a review]{2021Galax...9...66P}. One of these is the issue of lopsided satellite systems \citep{2016ApJ...830..121L, 2017ApJ...850..132P, 2020ApJ...898L..15B, 2021arXiv210412787W}. The M\,31 system was found to be highly lopsided, with a majority of its satellites residing on the side towards the Milky Way \citep{2013ApJ...766..120C}, an issue that in itself is very exceptional in a cosmological context (Kanehisa et al. in prep.). Interestingly, also the NGC\,253 system appears to be rather lopsided, with a preference of members to be mostly in the North of NGC\,253. The satellite distribution can be split in a slightly diagonal line to only contain only four out of 14 galaxies with measured distances that place them within 600\,kpc of NGC\,253 (bottom right half of Fig.~\ref{fig:onsky}). This asymmetry is also present in our additional candidates, of which four out of five are in the North, too.

\section{Conclusions}

In this paper we report the discovery of five dwarf spheroidal galaxies possible satellites of the bright late-type spiral NGC\,253 galaxy from a systematic visual inspection of the DESI Legacy Surveys images.

In Paper~I we gave a detailed description of the structure of the NGC\,253 group. We estimated the NGC\,253 group mass of about $8\times10^{11}$~M$_\sun$, virial and so-called turn-around radii, equal to $R_{200}=186$ and $R_\mathrm{ta}=706$~kpc respectively. Since then, the NGC\,253 galaxy group was revised with new members with known photometric distances and candidate member galaxies. The new list of the group is given in Table~\ref{tab:ngc253group}. The known number of satellite galaxies was increased from 9 to 15.

A group of galaxies around NGC\,7793 is located at the boundary of the system and just beginning their infall to the group. According to TRGB-distance measurements, the galaxies DDO\,226 and NGC\,59, although projected onto the group, are located outside its turn-around radius.

If confirmed as satellites then the new, more accurate map of the NGC\,253 group would illustrate the flat structure lying close to the plane of the Local Supercluster. The discovered dwarfs Do VIII and X would belong to this plane, while Do V, VI and IX should lead to its thickening.

Upon inclusion of the additional satellites with available distance measurements, the overall system's spatial distribution is now considerably wider than before, with an rms height of $\Delta_\mathrm{rms} = 87 \pm 6$\,kpc, and a minor-to-major axis ratio of $c/a = 0.35 \pm 0.05$. However, the distribution remains in an edge-on orientation and aligned with the Supergalactic plane. While more flattened than expected for isotropic systems, we find these values to be consistent with those of systems in the IllustrisTNG-100 hydrodynamical cosmological simulation. We also notice a lopsidedness of the satellite distribution, with a preference of 10 out of 14 objects with distance measurements to reside on one (approximately the northern) side of NGC\,253. 

Despite this progress, the final verdict on the presence of a correlated satellite structure in the NGC\,253 system is still out. The census of dwarfs remains incomplete. The candidates presented in this work require spectroscopic follow-up, and spatially-resolved photometric data, to accurately derive their distances and velocities. Without complete velocity information (beyond the only five objects with spectroscopic velocities available at the moment) the degree of kinematic coherence, the most important characteristic of other satellite planes, can not be assessed and tested against cosmological expectations.

\begin{acknowledgements}

DMD acknowledges the grant CNS2022-136017 funding by MICIU/AEI
/10.13039/501100011033 and the European Union NextGenerationEU/PRTR and
finantial support from the Severo Ochoa Grant CEX2021-001131-S funded by MCIN/AEI/
10.13039/501100011033. JS acknowledges financial support from project PID2022-138896NB-C53 and the Severo Ochoa grant CEX2021-001131-S funded by MCIN/AEI/ 10.13039/501100011033. DM and LM acknowledge support from the Russian Science Foundation grant \textnumero~24--12--00277, \url{https://rscf.ru/en/project/24-12-00277/}.
MSP acknowledges funding via a Leibniz-Junior Research Group (project number J94/2020).
We acknowledge the usage of the HyperLeda database\footnote{\url{http://leda.univ-lyon1.fr}}~\citep{2014A&A...570A..13M}.

This project used public archival data from the Dark Energy Survey. Funding for the DES Projects has been provided by the U.S. Department of Energy, the U.S. National Science Foundation, the Ministry of Science and Education of Spain, the Science and Technology FacilitiesCouncil of the United Kingdom, the Higher Education Funding Council for England, the National Center for Supercomputing Applications at the University of Illinois at Urbana-Champaign, the Kavli Institute of Cosmological Physics at the University of Chicago, the Center for Cosmology and Astro-Particle Physics at the Ohio State University, the Mitchell Institute for Fundamental Physics and Astronomy at Texas A\&M University, Financiadora de Estudos e Projetos, Funda{\c c}{\~a}o Carlos Chagas Filho de Amparo {\`a} Pesquisa do Estado do Rio de Janeiro, Conselho Nacional de Desenvolvimento Cient{\'i}fico e Tecnol{\'o}gico and the Minist{\'e}rio da Ci{\^e}ncia, Tecnologia e Inova{\c c}{\~a}o, the Deutsche Forschungsgemeinschaft, and the Collaborating Institutions in the Dark Energy Survey.

The Collaborating Institutions are Argonne National Laboratory, the University of California at Santa Cruz, the University of Cambridge, Centro de Investigaciones Energ{\'e}ticas, Medioambientales y Tecnol{\'o}gicas-Madrid, the University of Chicago, University College London, the DES-Brazil Consortium, the University of Edinburgh, the Eidgen{\"o}ssische Technische Hochschule (ETH) Z{\"u}rich,  Fermi National Accelerator Laboratory, the University of Illinois at Urbana-Champaign, the Institut de Ci{\`e}ncies de l'Espai (IEEC/CSIC), the Institut de F{\'i}sica d'Altes Energies, Lawrence Berkeley National Laboratory, the Ludwig-Maximilians Universit{\"a}t M{\"u}nchen and the associated Excellence Cluster Universe, the University of Michigan, the National Optical Astronomy Observatory, the University of Nottingham, The Ohio State University, the OzDES Membership Consortium, the University of Pennsylvania, the University of Portsmouth, SLAC National Accelerator Laboratory, Stanford University, the University of Sussex, and Texas A\&M University.

Based in part on observations at Cerro Tololo Inter-American Observatory, National Optical Astronomy Observatory, which is operated by the Association of Universities for Research in Astronomy (AURA) under a cooperative agreement with the National Science Foundation.

This work was partly done using GNU Astronomy Utilities (Gnuastro, ascl.net/1801.009) version 0.21. Work on Gnuastro has been funded by the Japanese Ministry of Education, Culture, Sports, Science, and Technology (MEXT) scholarship and its Grant-in-Aid for Scientific Research (21244012, 24253003), the European Research Council (ERC) advanced grant 339659-MUSICOS, the Spanish Ministry of Economy and Competitiveness (MINECO, grant number AYA2016-76219-P) and the NextGenerationEU grant through the Recovery and Resilience Facility project ICTS-MRR-2021-03-CEFCA.

This research has made use of the NASA/IPAC Infrared Science Archive, which is funded by the National Aeronautics and Space Administration and operated by the California Institute of Technology

\end{acknowledgements}

\bibliographystyle{aa} 
\bibliography{ref} 

\onecolumn 
 \begin{appendix}

\section{Update NGC 253 group}

\begin{table}[hbt!]
\centering
\caption{
Galaxies in $15\fdg5$ neighborhood around NGC~253
}
\begin{threeparttable}

\begin{tabular}{ll.@{$\pm$}00l.ll.00}
\hline\hline
Name &  
Type &
\multicolumn{2}{c}{$V_\mathrm{h}$} &
\multicolumn{1}{c}{$V_\mathrm{LG}$} &
&
\multicolumn{2}{c}{$D_\mathrm{TRGB}$} &
&
\multicolumn{1}{c}{$\Theta_\mathrm{N253}$} &
\multicolumn{1}{c}{$R_\perp$} &
\multicolumn{1}{c}{$R_\mathrm{3D}$} \\
\cline{3-5}\cline{7-8}\cline{11-12}
&
&
\multicolumn{3}{c}{(km\,s$^{-1}$)} &
&
\multicolumn{2}{c}{(Mpc)} &
&
\multicolumn{1}{c}{(deg)} &
\multicolumn{2}{c}{(kpc)} \\
\hline

NGC 253       & SABc &   261 &    5 &  294 & \tnote{a} & 3.70 & $_{-0.03}^{+0.03} $ & \tnote{$\dagger$}  &       &      &      \\
Scl-MM-Dw2    & dSph &\multicolumn{4}{c}{}             & 3.53 & $_{-0.11}^{+0.12} $ & \tnote{$\ddagger$} &  0.82 &   53 &  174 \\
Scl-MM-Dw1    & dSph &\multicolumn{4}{c}{}             & 3.52 & $_{-0.50}^{+0.58} $ & \tnote{$\ddagger$} &  1.10 &   71 &  195 \\
Do II         & dSph &\multicolumn{4}{c}{}             & 3.47 & $_{-0.28}^{+0.15} $ & \tnote{$\ddagger$} &  1.34 &   87 &  246 \\
Scl-MM-Dw4    & dSph &\multicolumn{4}{c}{}             & 4.11 & $_{-0.33}^{+0.17} $ & \tnote{$\ddagger$} &  1.42 &   92 &  424 \\
Scl-MM-Dw5    & dSph &\multicolumn{4}{c}{}             & 3.89 & $_{-0.26}^{+0.18} $ & \tnote{$\ddagger$} &  1.57 &  101 &  219 \\
Do IX         &      &\multicolumn{4}{c}{}             & \multicolumn{3}{c}{}                            &  1.88 &  122 &      \\
LV J0055-2310 & dIr  &   250 &    5 &  288 & \tnote{a} & 3.62 & $_{-0.18}^{+0.19} $ & \tnote{$\flat$}    &  2.72 &  175 &  190 \\
Do IV         & dSph &\multicolumn{4}{c}{}             & 3.94 & $_{-0.56}^{+0.65} $ & \tnote{$\sharp$}   &  3.61 &  233 &  344 \\
Do V          &      &\multicolumn{4}{c}{}             & \multicolumn{3}{c}{}                            &  3.89 &  251 &      \\
dw0036$-$2828 & dIr  &\multicolumn{4}{c}{}             & 3.77 & $_{-0.33}^{+0.36} $ & \tnote{$\sharp$}   &  4.02 &  260 &  271 \\
DDO 6         & dIr  & 291.8 &  1.6 &  344 & \tnote{b} & 3.44 & $_{-0.15}^{+0.13} $ & \tnote{$\dagger$}  &  4.31 &  278 &  375 \\
Do VI         &      &\multicolumn{4}{c}{}             & \multicolumn{3}{c}{}                            &  4.39 &  283 &      \\
NGC 247       & SABd &   153 &    5 &  208 & \tnote{a} & 3.72 & $_{-0.03}^{+0.03} $ & \tnote{$\dagger$}  &  4.53 &  293 &  294 \\
Do VII        &      &\multicolumn{4}{c}{}             & \multicolumn{3}{c}{}                            &  4.74 &  306 &      \\
Do VIII       &      &\multicolumn{4}{c}{}             & \multicolumn{3}{c}{}                            &  5.00 &  323 &      \\
Do III        & dSph &\multicolumn{4}{c}{}             & 3.37 & $_{-0.17}^{+0.18} $ & \tnote{$\sharp$}   &  5.31 &  342 &  461 \\
Sc 22         & dSph &\multicolumn{4}{c}{}             & 4.29 & $_{-0.06}^{+0.08} $ & \tnote{$\dagger$}  &  5.40 &  349 &  697 \\
ESO 540-032   & dIr  & 227.7 &  0.9 &  285 & \tnote{c} & 3.63 & $_{-0.05}^{+0.05} $ & \tnote{$\dagger$}  &  5.42 &  350 &  353 \\
KDG 2         & dTr  &   224 &    3 &  290 & \tnote{c} & 3.56 & $_{-0.07}^{+0.07} $ & \tnote{$\dagger$}  &  7.23 &  466 &  477 \\[5pt]

PGC 704814    & dIr  &   270 &   89 &  299 & \tnote{d} & 3.66 & $_{-0.16}^{+0.17} $ & \tnote{$\flat$}    & 12.39 &  798 &  795 \\
ESO 349-031   & dIr  &   220 &    5 &  229 & \tnote{a} & 3.21 & $_{-0.06}^{+0.06} $ & \tnote{$\dagger$}  & 12.59 &  811 &  901 \\
NGC 7793      & SAd  &   224 &    5 &  247 & \tnote{a} & 3.63 & $_{-0.13}^{+0.14} $ & \tnote{$\dagger$}  & 13.09 &  843 &  838 \\[5pt]

DDO 226       & Ir   & 358.6 &  1.6 &  409 & \tnote{b} & 4.92 & $_{-0.26}^{+0.30} $ & \tnote{$\dagger$}  &  3.21 &  207 & 1245 \\
Sculptor SR   & Tr   &\multicolumn{4}{c}{}             & 19   &                     & \tnote{$\sharp$}   &  3.99 &  257 &      \\
NGC 59        & dEem &   368 &    5 &  438 & \tnote{a} & 4.90 & $_{-0.07}^{+0.07} $ & \tnote{$\dagger$}  &  8.32 &  537 & 1349 \\
UGCA 442      & SBm  &   267 &    5 &  299 & \tnote{a} & 4.37 & $_{-0.16}^{+0.14} $ & \tnote{$\dagger$}  & 15.48 &  996 & 1271 \\

\hline\hline
\end{tabular}

\begin{tablenotes}
\textbf{The columns:}
(1) galaxy name;
(2) morphological type according to the HyperLeda~\citep{2014A&A...570A..13M} and the Local
Volume~\citep{2012AstBu..67..115K} databases;
(3) $V_\mathrm{h}$ --- heliocentric velocity in km\,s$^{-1}$ with its error;
(4) $V_\mathrm{LG}$ --- radial velocity with respect to the LG centroid \citep{1996AJ....111..794K};
(5) $D_\mathrm{TRGB}$ --- TRGB–distance in Mpc with a corresponding error;
(6) $\Theta_\mathrm{N253}$ --- angular separation from NGC 253 in degrees;
(7) $R_\perp$ and (8) $R_\mathrm{3D}$ --- projected and spatial distances to NGC 253 in kpc, correspondingly.

\textbf{Velocity references:}
\item[a] \citet{2017MNRAS.472.4832W},
\item[b] \citet{2008MNRAS.386.1667B},
\item[c] \citet{2005AJ....130.2058B},
\item[d] \citet{2003astro.ph..6581C}.

\textbf{TRGB distance references:}
\item[$\dagger$] \citet{2021AJ....162...80A},
\item[$\ddagger$] \citet{2022ApJ...926...77M},
\item[$\flat$] \citet{2021AJ....161..205K},
\item[$\sharp$] \citet{Mutlu-Pakdil2024}.
\end{tablenotes}

\end{threeparttable}

\label{tab:ngc253group}

\end{table}

 \end{appendix}
 \end{document}